\documentclass[journal=nalefd,manuscript=letter,layout=twocolumn,maxauthors=50]{achemso}
\usepackage{dcolumn}
\setkeys{acs}{articletitle = true,etalmode=truncate,maxauthors=10}
\usepackage{amsmath}
\usepackage{bm}
\usepackage{rotating}
\usepackage{multirow}
\usepackage[table]{xcolor}
\usepackage[english]{babel}

\def\ve{{\varepsilon}}

\def\bk{{\bf k}}

\def\>{\rangle}
\def\<{\langle}

\title{Two-dimensional plasmonic polarons in $\textbf{\emph{n}}$-doped monolayer MoS$_2$ }

\author{Fabio Caruso}
\email{caruso@physik.uni-kiel.de}
\affiliation{Institut f\"ur Physik and IRIS Adlershof, Humboldt-Universit\"at zu Berlin, Berlin, Germany} 
\alsoaffiliation{  Institut f\"ur Theoretische Physik und Astrophysik,  Christian-Albrechts-Universit\"at zu Kiel, D-24098 Kiel, Germany}
\author{Patrick Amsalem}
\affiliation{Institut f\"ur Physik and IRIS Adlershof, Humboldt-Universit\"at zu Berlin, Berlin, Germany} 
\author{Jie Ma}
\affiliation{Institut f\"ur Physik and IRIS Adlershof, Humboldt-Universit\"at zu Berlin, Berlin, Germany} 
\author{Areej Aljarb}
\affiliation{Physical Sciences and Engineering Division, King Abdullah University of Science and Technology, Thuwal, 23955-6900, Kingdom of Saudi Arabia}
\author{Thorsten Schultz}
\affiliation{Institut f\"ur Physik and IRIS Adlershof, Humboldt-Universit\"at zu Berlin, Berlin, Germany} 
\alsoaffiliation{Helmholtz-Zentrum f\"ur Materialien und Energie GmbH, Berlin, Germany}
\author{Marios Zacharias}
\affiliation{Department of Mechanical and Materials Science
  Engineering, Cyprus University of Technology, P.O. Box 50329, 3603
  Limassol, Cyprus}
\author{Vincent Tung}
\affiliation{Physical Sciences and Engineering Division, King Abdullah University of Science and Technology, Thuwal, 23955-6900, Kingdom of Saudi Arabia}
\author{Norbert Koch}
\affiliation{Institut f\"ur Physik and IRIS Adlershof, Humboldt-Universit\"at zu Berlin, Berlin, Germany} 
\alsoaffiliation{Helmholtz-Zentrum f\"ur Materialien und Energie GmbH, Berlin, Germany}
\author{Claudia Draxl}
\affiliation{Institut f\"ur Physik and IRIS Adlershof, Humboldt-Universit\"at zu Berlin, Berlin, Germany}

\begin{document}

\begin{abstract}
We report experimental and theoretical evidence of strong
electron-plasmon interaction in $n$-doped single-layer MoS$_2$. 
Angle-resolved photoemission spectroscopy (ARPES) measurements reveal the emergence of distinctive 
signatures of polaronic coupling in the electron spectral function. 
Calculations based on many-body perturbation theory illustrate that 
electronic coupling to two-dimensional (2D) carrier plasmons 
provides an exhaustive explanation of the experimental spectral features 
and their energies. 
These results constitute compelling evidence of the 
formation of plasmon-induced polaronic quasiparticles, 
suggesting that highly-doped transition-metal 
dichalcogenides may provide a new platform to explore 
strong-coupling phenomena between electrons and plasmons in 2D. 
\end{abstract}

\maketitle

The interplay of charge confinement, reduced dielectric screening, 
and strong light-matter coupling in few-layer semiconducting 
transition-metal dichalcogenides (TMDCs) underpins a vast 
spectrum of emergent many-body effects, including the formation of 
excitons \cite{Mak2010,Qiu2013,Molina2016,hong2020}, trions \cite{mak_tightly_2013}, 
polarons \cite{Kang2018,Garcia-Goiricelaya}, polaritons 
\cite{Zheng2017,stuhrenberg2018,Asger2018,Liu2018,Cuadra2018}, 
and superconductivity \cite{costanzo_gate-induced_2016}. 
At variance with three-dimensional solids, TMDCs further allows for 
unprecedented opportunities to tailor these phenomena
via cavity embedding \cite{Latini2019},  circular dichroism \cite{cao_valley-selective_2012,beyer_80_2019}, 
gating \cite{costanzo_gate-induced_2016}, nano-structuring \cite{kang_high-mobility_2015,Cuadra2018}, 
substrate engineering \cite{briggs_atomically_2020}, and doping \cite{Kang2018}. 

Highly-tunable carrier densities are particularly 
desirable for the study of many-body interactions in TMDCs, 
as they may enable control of plasmons 
(collective excitation of the electron density) and 
polarons (electrons dressed by a phonon cloud) \cite{Moser2013prl,Baumberger2016}.
Polarons typically 
arise in semiconductors and insulators as a result 
of strong coupling to longitudinal-optical (LO) 
vibrational modes \cite{Moser2013prl,Janotti,Baumberger2016,Verdi2017}. 
Polarons may lead to charge trapping \cite{Janotti,Moses2016}
and to a renormalization of the band effective masses \cite{Verdi2017}. 
The relevance of such phenomena for the opto-electronic 
properties of solids, alongside with the recent discovery of polarons in the photoemission 
spectrum of doped oxides \cite{Moser2013prl,Baumberger2016} and 2D materials \cite{Kang2018,Chen2018},
has reignited theoretical and experimental research on polaronic quasiparticles 
\cite{Janotti,Moses2016,Verdi2017,EuO,Caruso/PRB/2018,Garcia-Goiricelaya,Sio/2019,Sio20192}.

In close analogy to phonon-induced polarons, 
the formation of polaronic quasiparticles may also be stimulated by 
the coupling to extrinsic plasmons in highly-doped semiconductors \cite{EuO},  
leading to distinctive satellite features
in angle-resolved photoemission spectroscopy (ARPES). 
The experimental observation of plasmonic polarons in three-dimensional 
solids is hindered by rather low electron-plasmon coupling 
strengths ($\lambda \le 0.5$ \cite{EuO}), and by the difficulty to achieve 
degenerate doping concentrations (a prerequisite 
for the excitation of plasmons) while simultaneously retaining 
crystallinity of the sample. 
{Single-layer TMDCs circumvent this problem since 
highly-tunable carrier densities can be realized. 
In particular, some of us have recently demonstrated 
that the carrier population in the conduction band of 
monolayer MoS$_2$ and WS$_2$ can be tailored by stimulating the   
formation of chalcogen vacancies via repeated 
annealing cycles \cite{Amsalem2020}.  
Additionally, we show here that the 2D confinement 
of plasmons and charge carriers provide suitable 
conditions for the establishment of a strong-coupling regime, leading to the 
emergence of plasmonic-polaron quasiparticles. 
}

In this manuscript, we report the observation of plasmon-induced 
polarons in $n$-doped monolayer MoS$_2$.
ARPES measurements of the conduction band reveals 
distinctive fingerprints of bosonic satellites due to 
electron-boson coupling. 
{The satellite energy, ranging 
between 130 and 200~meV, is much larger than the characteristic 
phonon energies ($\hbar\omega < 60$~meV), and 
depends pronouncedly on the carrier density. 
These characteristics are incompatible with 
a coupling mechanism due to LO phonons.} 
By explicitly accounting for the effects of electronic 
coupling to 2D plasmons in a first-principles many-body framework, 
our calculations of the spectral function identify electron-plasmon interactions 
as the origin of the polaronic spectral features observed in experiments. 
{These findings provide compelling evidence of the emergence 
of 2D plasmonic polarons in $n$-doped monolayer MoS$_2$, 
and they demonstrate the establishment of a strong-coupling 
regime resulting from the confinement of carriers and plasmons in 2D. }

The measured samples consist of a chemical-vapor-deposition (CVD) 
grown fully-closed MoS$_2$ monolayer film with azimuthal disorder 
on a sapphire substrate \cite{Aljarb}. 
Top mechanical clamping of the sample enabled electrical contact of the monolayer 
to ground. Sample charging during ARPES measurements was avoided upon doping 
of the monolayer as described thereafter. 
Degenerate $n$-doping was achieved by repeated {\it in-situ} 
vacuum annealing cycles (with final annealing for 12 hours at 850~K), 
which is known to effectively $n$-dope MoS$_2$ monolayer \cite{Baugher}. 
{The Fermi-level position 
(corresponding to 0 eV binding energy) and the instrumental 
broadening were determined by fitting the Fermi edge of a 
gold polycrystal using a broadened Fermi-Dirac function. 
The estimate doping concentrations  $n_1= 2.8\cdot 10^{13}$,  $n_2 = 4\cdot 10^{13}$,
and $n_3 = 5\cdot 10^{13}$~cm$^{-2}$ (referred to simply as $n_1$, $n_2$, $n_3$ in the following),} result 
from the formation of sulfur vacancies either by direct 
desorption of sulfur atoms or by desorption of substitutional 
oxygen present at sulfur-vacancy sites, naturally present in 
CVD grown samples \cite{Schuler,Liu}. 
Our calculations indicate that this concentration of S vacancies 
leaves the band structure unchanged (Supplementary Note~5 and Fig.~S4 in the Supporting Information (SI)).  
The photoemission measurements on $n_1$ and $n_2$ were performed in an analysis chamber (base pressure $2 \cdot 10^{-10}$~mbar) 
equipped with a Specs Phoibos 100 hemispherical electron analyzer and using the HeI 
radiation provided by monochromated helium discharge lamp (consisting of a HIS-13 
lamp mounted on a VUV5046 UV-monochromator). 
These measurements were performed at room temperature. 
The overall energy resolution amounted to 117~meV 
(65~meV instrumental energy resolution) as determined 
from the Fermi edge of a polycrystalline gold sample 
and the angular resolution was about $\pm2$ degrees. 
{ The ARPES measurements on $n_3$ were performed at 230~K
using a DA30-L hemispherical analyzer with an angular 
resolution of ca. 0.3~$^{\circ}$  and an overall resolution of 
100 meV (65 meV instrumental energy resolution). }
The energy distribution curves were measured every 
two degrees by rotating the manipulator about the polar axis.
Despite photoemission averaging over all azimuthal orientations 
as exhibited by the MoS$_2$ grains, the low dimensionality of the 
system promotes the observation in the ARPES spectra of 
well-resolved electronic band dispersion 
along the high-symmetry directions $\Gamma$-K 
and $\Gamma$-M \cite{park_electronic_2019}.

\begin{figure}[t]
\begin{center}
\includegraphics[width=0.48\textwidth]{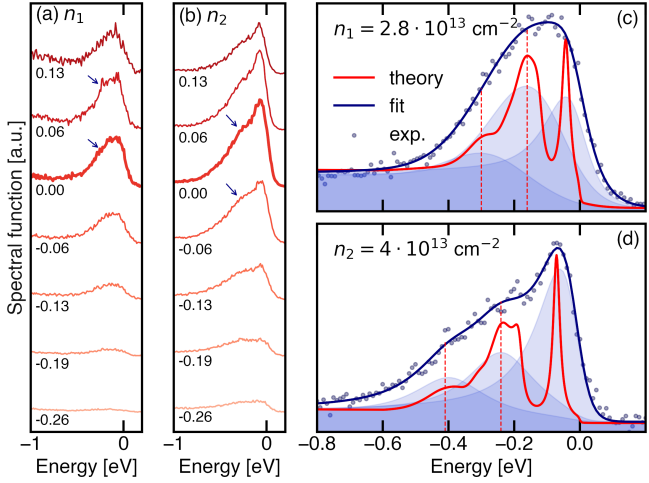}
\caption{
(a) Spectral signatures of plasmonic polarons in 
the ARPES spectral function of $n$-doped 
monolayer MoS$_2$ for the carrier densities (a) $n_1= 2.8\cdot 10^{13}$ and (b)  $n_2 = 4\cdot 10^{13}$~cm$^{-2}$.
Labels indicate the photo-electron crystal momentum relative to 
the K point ($\Delta{\bf k} = {\bf k} - {\bf k}_{\rm K}$) along 
K-$\Gamma$ in units of \AA$^{-1}$, and satellite features are indicated by arrows.
Energies are relative to the Fermi level. 
ARPES measurements (dots) and first-principles 
calculations (red) of the spectral function at
(c) K for $n_1$ and  (d) $n_2$. 
Vertical dashed lines mark the positions of the first 
and second satellite peaks in the calculated spectra. 
The fit (dark blue)  of the experimental spectra 
and its spectral decomposition as Gaussian 
lineshapes (shaded curves) are guides to the eye.  \label{fig:spec_vs_n}
}
\end{center}
\end{figure}

Figures~\ref{fig:spec_vs_n}~(a) and (b) report ARPES spectral functions 
of monolayer MoS$_2$ for crystal momenta in the vicinity of the  high-symmetry 
point K for doping concentrations $n_1$ and $n_2$, respectively, 
whereas the full measured spectral function for the valence band 
is shown in Fig.~S5 in the SI.
The on-curve labels indicate the photo-electron crystal momentum 
$\Delta{\bf k} = {\bf k} - {\bf k}_{\rm K}$, relative to the K point 
in units of \AA$^{-1}$. Energies are referenced to the Fermi level that is 
located 39 and 56~meV above the conduction-band minimum (CBm), respectively.
At K and neighboring crystal momenta, 
the spectral functions are characterized by a 
quasiparticle peak close to the Fermi level. 
The peak position corresponds to the binding energy of 
quasi-electrons around the CBm. 
At binding energies larger than those of quasiparticle excitations,
the spectral function for carrier concentration $n_1$ ($n_2$)
exhibits a shoulder-like structure red-shifted by 130~meV (170~meV) from 
the quasiparticle peak and extending up to 0.6~eV below the Fermi level.
These spectral features, indicated by arrows in 
Figs.~\ref{fig:spec_vs_n}~(a) and (b), become 
more pronounced at higher carrier densities. 

A more detailed view of these spectral signatures is 
given in Figs.~\ref{fig:spec_vs_n}~(c) and (d), which report 
the spectral functions for $n_1$ and $n_2$
(corresponding to the thick lines in Figs.~\ref{fig:spec_vs_n}~(a) and (b)),
{whereas measurements for a third carrier concentration ($n_3$)
are reported in Fig.~S7 (see SI).}
For $n_2$, the spectral function exhibits a satellite structure at 170~meV and
a less-pronounced secondary structure at 330~meV below the quasiparticle peak. 
These features become more pronounced 
for momenta close to the K point at $\Delta {\bf k} = 0.06$~\AA$^{-1}$ (Fig.~S6 in the SI). 
{To better resolve the underlying spectral features that give rise to 
the satellite, in Fig.~S8 we apply the prescription of 
Ref.~\cite{ANDERSON19901700} and de-convolute the experimental 
spectra by a Gaussian function to reduce the spectral broadening 
due to finite resolution and lifetime effects.}
These spectral features closely resemble the characteristic spectral 
fingerprints of electron-boson interaction in photoemission 
spectroscopy. Specifically:  
(i)  they have non-vanishing intensity within the gap 
and, therefore, may not be attributed to the emission of a photo-electron; 
{(ii) they are roughly spaced from the quasiparticle peak by multiples 
of $\hbar\Omega=$130, 170, and 200~meV for $n_1$, $n_2$, and $n_3$, respectively, 
as illustrated by the spectral-function decomposition 
in terms of Gaussian lineshapes shown in Figs.~\ref{fig:spec_vs_n}~(c) and (d) and Fig.~S7.}
These points suggest that carriers in the conduction band 
may be subject to polaronic coupling to bosonic excitations -- such 
as, e.g., polar phonons or 2D carrier plasmons 
 --  and that the first and second satellites may stem from 
the excitation of one and two bosons with an effective 
energy $\hbar\Omega$ alongside with the creation of a photo-hole.

The comparison between the boson energy $\hbar\Omega$ and 
the energy of optical phonons in monolayer MoS$_2$, that ranges between 
between 30 and 60~meV (see Fig.~S9 of the SI), 
allow us to promptly exclude the Fr\"ohlich interaction 
between electrons and LO phonons as possible 
coupling mechanism responsible for the satellite formation.
Recent experimental and theoretical investigations of {\it bulk }
MoS$_2$ \cite{Kang2018,Garcia-Goiricelaya} have indeed revealed 
the emergence of Holstein polarons in ARPES. 
However, their spectral signatures occur at 
30-40~meV below the Fermi level,  energies that 
are too small to possibly explain the polaronic 
spectral features of Fig.~\ref{fig:spec_vs_n}. 
In analogy to polarons in oxides \cite{Sio/2019}, 
strong coupling between electrons and plasmons may also
trigger the formation of plasmon-induced polaronic 
quasiparticles ({\it plasmonic polarons})
in $n$-doped semiconductors and insulators.  
Plasmonic polarons result from the simultaneous excitation of a 
plasmon and a hole, and they manifest themselves in ARPES under 
the form of photoemission satellites. 
{ The satellite energy observed in our measurements 
depends pronouncedly on the carrier concentration (Fig.~S10 of SI), 
suggesting a coupling mechanism related to electronic degrees of freedom 
which is compatible with this picture.  }
To inspect whether a bosonic coupling mechanism induced by 
carrier plasmons may account for these phenomena, we proceed 
to investigate the plasmon dispersion in doped MoS$_2$. 

\begin{figure}[t]
\begin{center}
\includegraphics[width=0.48\textwidth]{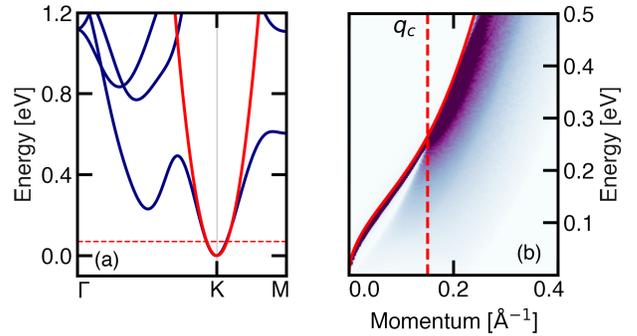}
\caption{(a) Conduction band of {monolayer} MoS$_2$ obtained from density-functional theory (blue), 
and dispersion of a 2D homogeneous electron gas with effective mass $m_{\rm DFT}^*=0.49$ (red). 
The dashed horizontal lines mark the position of the Fermi level $\epsilon_{\rm F}$
for the carrier concentration $n_2$. 
Energies are referenced to the conduction-band minimum. 
(b) Loss function and plasmon dispersion of $n$-doped MoS$_2$ 
for the carrier concentration $n_2$.  
The vertical dashed line marks the critical momentum 
cutoff ${\bf q}_c$ corresponding to the onset of Landau 
damping for 2D plasmons.
The continuous line illustrates the plasmon dispersion.\label{fig:pl_disp} 
}
\end{center}
\end{figure}

For concentrations of $n$-type carriers larger than $1\cdot10^{13}$~cm$^{-2}$,
MoS$_2$ undergoes a metal-insulator transition \cite{Mott/RMP/1968},
characterized by the onset of metallic transport properties \cite{MIT/mos2/2013} and by 
the partial filling of the K valley in the conduction band. 
The conduction-band dispersion {of monolayer  MoS$_2$} as obtained from density-functional theory 
calculations is reported in Fig.~\ref{fig:pl_disp}~(a).
The Fermi energy of the doped system depends linearly 
on the carrier concentration $n$ via the expression 
$\epsilon_{\rm F} = \pi \hbar^2 n / m^* N_{\rm v} - E_{\rm CBm}$,
with $N_{\rm v}=2$ being the K-valley degeneracy, 
and $E_{\rm CBm}$ the energy of the CBm.
{For the experimentally determined Fermi energies $\epsilon_{\rm F}=39$, $56$, and $70$~meV, 
this expression allows one to determine the carrier concentrations $n_1$, $n_2$, and 
$n_3$ respectively.} 
For $n_2$, the Fermi energy is marked by the horizontal dashed line in Fig.~\ref{fig:pl_disp}~(a). 
Upon interaction with light, extrinsic carriers in the conduction band 
can be collectively excited leading to well-defined 2D carrier plasmon resonances. 
The energy-momentum dispersion relation of 2D plasmons in MoS$_2$ is here approximated by  
the loss function $L(\omega) = {\rm Im}~[\epsilon({\bf q,\omega})]^{-1}$
of a 2D homogeneous electron gas (2HEG) \cite{Caruso2016EPJB,Caruso/PRB2/2018}. 
All calculations have been conducted using the experimental effective mass 
$m_{\rm exp}^*=0.9~m_{\rm e}$ (with $m_{\rm e}$ being the bare electron mass), which has been 
obtained by ARPES measurements of the conduction band of monolayer MoS$_2$ single-crystals 
(see Supplementary Note~3 and Fig.~S1). The influence of $m^*$ on the calculations and  
on electron-plasmon interaction is discussed in Supplementary Note~4 and Fig.~S2-S3. 
We further explicitly account for the extrinsic dielectric environment 
induced by the semi-infinite sapphire (Al$_2$O$_3$) substrate 
by introducing the dielectric constant 
$\epsilon^{\rm S}_\infty = ( 1 + \epsilon^{\rm Al_2O_3}_\infty ) / 2 = 6.3$, 
with $\epsilon^{\rm Al_2O_3}_\infty \simeq 11.6$
being the high-frequency dielectric constant of 
bulk sapphire \cite{Fontanella/1974}.

The loss function, illustrated in Fig.~\ref{fig:pl_disp}~(b)
for the carrier concentration $n_2$, exhibits pronounced plasmon 
peaks, with plasmon energies $\hbar\omega_{\rm pl}({\bf q})$ ranging from 0 to 0.24~eV. 
Only plasmons with momenta $| {\bf q}|  < { q}_{\rm c}$  
can be excited, where ${ q}_{\rm c} \simeq 0.15~$\AA$^{-1}$ (marked by a 
dashed vertical line  in Fig.~\ref{fig:pl_disp}~(b)) is the critical momentum for 
the onset of Landau damping, namely, the decay of plasmons upon excitation of 
electron-hole pairs. 
The plasmon dispersion obtained from the loss function
compares well with the analytical result 
for homogeneous 2D metals
{$\omega^{\rm pl}_{\bf q} =   F(q)\sqrt{ { 2\pi n q}/ {m^* \epsilon^{\rm S}_\infty} }$ \cite{Giuliani/2005}, 
illustrated in Fig.~\ref{fig:pl_disp}~(b) as a continuous line. Here, 
$F(q) = (1+q/2\kappa) [(1+q/2\kappa)^{-1}+q^3 N_v/(4\pi n \kappa )]^{\frac{1}{2}}$ 
and $\kappa = 2 m^*/ \epsilon^{\rm S}_\infty$.}

{In systems characterized by electronic coupling to weakly-dispersive bosonic modes 
(as, e.g., optical phonons or 3D plasmons) the energy of polaronic satellites
can be directly related to the boson energy. Therefore, at a first sight 
it might seem surprising that the coupling to 2D plasmons 
could give rise to a satellite peak with a well-defined energy, 
since the 2D plasmon dispersion continuously spans the energy 
range between 0 and 240~meV (Fig.~\ref{fig:pl_disp}~(b)).
To clarify this aspect, we show in the following 
that the satellite energy is related to the average 
energy of 2D plasmons in the Brillouin zone. 
This finding leads us to the interpretation of
the measured ARPES satellite as a result of the 
collective coupling to all 2D plasmons in the system, 
rather than to a single bosonic excitation.  
To estimate the average energy of 2D plasmons, we consider the 
density of states of plasmonic excitations:
}
\begin{equation}\label{eq:pl_DOS}
J(\omega) = {\Omega_{\rm BZ}}^{-1}\int_{\Omega_{\rm BZ}} 
{d{\bf q}} \delta(\omega-\omega^{\rm pl}_{\bf q}) =\kappa \omega^3 \theta(\omega_{\rm c} -\omega)
\end{equation} 
where  $\kappa = ( m^* \epsilon^{\rm S}_\infty )^2(\pi{\Omega_{\rm BZ }n^2)^{-1}}$ 
and $\omega_{\rm c} = \omega^{\rm pl}_{{\bf q}_{\rm c}}$ is the plasmon energy 
at the critical momentum ${\bf q}_{\rm c}$. 
{This result may be promptly verified by analytical integration of Eq.~\eqref{eq:pl_DOS}, as
illustrated in the SI. For $n_1$, $n_2$, and $n_3$, the density of states $J(\omega)$ is illustrated in Fig.~S11.}
From Eq.~\eqref{eq:pl_DOS}, we can estimate the average plasmon energy as 
$\hbar \overline \omega = \hbar { \int J(\omega) \omega d\omega}[{ \int J (\omega) d\omega}]^{-1} 
= {4\hbar \omega_{\rm c} }/{ 5 }$ and the standard deviation $\sigma = \hbar ( \overline{\omega^2} 
- \overline\omega^2  )^{1/2}\simeq 0.45 \hbar\omega_{\rm c}$. 
For $n_2$, the value $\hbar \omega_c\simeq 0.24$~eV yields 
$\hbar\overline \omega = 190$~meV which agrees well with 
the effective boson energy $\hbar\Omega = 170$~meV derived from the satellite energy. 
Similarly, the spread of the plasmon energies in the Brillouin zone is 
quantified by $2\sigma = 80$~meV, which is in 
good agreement with the half width at half maximum of the 
satellite peak $\Delta w = 90$~meV, which we extract from 
Gaussian decomposition of the experimental spectra. 
{This analysis suggests that the observation of a 
well-defined satellite in ARPES is compatible 
with a bosonic coupling mechanism induced by 
2D carrier plasmons. In particular, the satellite 
energy can be related to the average energy 
$\hbar\overline \omega$ of 2D plasmons, whereas the 
satellite linewidth reflects the spread of plasmon energies as
quantified by the standard deviation $\sigma$.}

To quantitatively demonstrate the influence of the 
electron-plasmon interaction on ARPES measurements, we conducted 
first-principles calculations of the spectral function based on the
cumulant expansion approach \cite{Langreth1970,Ferdi1996,guzzo/2011}, the state of the art for 
spectral-function calculations of coupled electron-boson systems \cite{Caruso2018book}. The cumulant spectral function can be 
expressed as \cite{Verdi2017}: 
  \begin{align} \label{eq.cumulant}
  A(\bk,\omega)  = \sum_n e^{A^{\rm S}_{n\bk}(\omega)\ast} A_{n\bk}^{\rm QP} (\omega)
  \end{align}
where $\ast$ denotes convolution over frequency, and 
$A_{n\bk}^{\rm{QP}}(\omega)= 2\pi^{-1}{\rm Im\,} [\hbar \omega-\ve_{n\bk}- \Sigma_{n\bk}(\ve_{n\bk})]^{-1}$
is the quasiparticle contribution to the spectral function. $\ve_{n\bk}$ is the single-particle energy, 
and  $\Sigma$ is the electron self-energy due to the electron-plasmon interaction, 
for which an explicit expression is given in the SI.
Dynamical correlations due to the electron-plasmon interaction are 
accounted for by the satellite function $A^{\rm S}_{n\bk}$:
 \begin{align} 
  A^{\rm S}(\omega)
  &=  \frac{\beta(\omega) - \!\beta(\varepsilon/\hbar) -
  \!(\omega-\varepsilon/\hbar)\!\left.
  \displaystyle({\partial \beta}/{\partial \omega})
    \right|_{\varepsilon/\hbar}}
  {(\hbar\omega-\varepsilon)^2}, 
 \end{align}
with $\beta(\omega) = \frac{1}{\pi}{\rm Im}\,\varSigma(\varepsilon/\hbar-\omega)
\theta({\mu}/\hbar-\omega)$,  and the dependence on $n$ and ${\bf k}$ has been omitted. 
Computational details are reported in Supplementary Note 6. 

\begin{figure}[!t]
\includegraphics[width=0.48\textwidth]{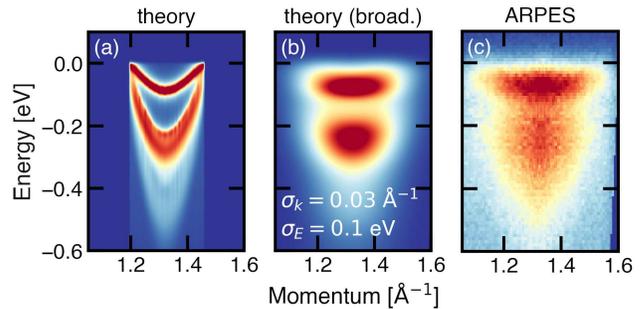}
\caption{ (a) First-principles calculations of the angle-resolved 
spectral function of MoS$_2$ at K for a carrier concentration 
$n_3 = 5\cdot 10^{13}$~cm$^{-2}$. 
(b) Influence of finite energy and momentum resolutions on the 
angle-resolved spectral function. 
(c) ARPES measurements.  \label{fig:disp}}
\end{figure}

The cumulant spectral function of $n$-doped MoS$_2$ is illustrated 
in Figs.~\ref{fig:spec_vs_n}~(c)-(d) as a red line for $n_1$ and $n_2$, respectively. 
{The spectral function for a broader range of carrier densities is 
illustrated in Fig.~S10~(b)}. 
{Remarkably, the increase of the satellite energy with carrier concentration 
is in excellent agreement with the calculated trend (see Fig.~S10~(a)).} 
The experimental background signal has been added to 
theoretical data to facilitate the comparison. 
The calculated spectral function exhibits two distinct satellite resonances 
at energies in excellent agreement with the ARPES spectral features.
The Taylor expansion of Eq.~\eqref{eq.cumulant} up to second order 
in $A^{\rm S}$ promptly reveals the physical origin of these spectral 
features: beside the quasiparticle peak, arising from $A^{\rm QP}$ alone, 
the second term in the expansion stems from the convolution 
 $A^{\rm S} \ast A^{\rm QP}$, and its intensity may be attributed 
to the rate of photoemission processes resulting from the coupled 
excitation of a photo-hole and a plasmon. 
Similarly, subsequent arguments in the Taylor expansion 
can be related to multiple plasmon excitations.
This allows us to relate subsequent satellites to photoemission processes 
resulting from the simultaneous excitation a photo-hole and emission of plasmons. 
The quantitative agreement between experiment and theory is 
further illustrated in Fig.~\ref{fig:disp}, where the 
full calculated and measured angle-resolved spectral functions
are compared for the carrier concentration $n_3=5\cdot10^{13}$~cm$^{-2}$.  
Finite experimental resolution for energy (momentum)
is explicitly accounted for in Fig.~\ref{fig:disp}~(b)
by convolution with a Gaussian function with 
full-width at half maximum of 100~meV (0.03~\AA$^{-1}$). 
The calculated satellite structure is sharper 
as compared to the measurements, since our calculations do not explicitly 
account for finite lifetime effects in the plasmon dispersion arising 
from the plasmon-phonon scattering. 
Scattering between plasmons and phonons can indeed lower the plasmon 
lifetime, introducing additional broadening of the plasmon 
satellite \cite{Caruso/PRB2/2018}.

In conclusion, we reported the observation 
of 2D plasmonic polarons in $n$-doped monolayer MoS$_2$
at degenerate doping concentrations. 
The emergence of distinctive signatures of polaronic satellites 
in ARPES experiments provides compelling evidence of the strong 
coupling between 2D carrier plasmons and extrinsic carriers, 
and it is corroborated by first-principles calculations of 
the electron-plasmon interaction. 
This finding indicates that, at sufficiently large 
doping concentrations, the quantum confinement of 
holes and plasmons in a 2D semiconductor 
triggers the onset of a strong-coupling regime, whereby 
electron-plasmon coupling results in the formation of 
2D plasmonic polarons. 

\begin{acknowledgement}
This work was funded by the Deutsche Forschungsgemeinschaft (DFG) 
-- Projektnummer 182087777 -- SFB 951.
FC acknowledges useful discussions with Pasquale Pavone. 
AA and VT are indebted to the support from the KAUST
Office of Sponsored Research (OSR) under Award No:
OSR-2018-CARF/CCF-3079. 
MZ acknowledges support by the Research Promotion Foundation of the Republic of Cyprus under Award:
NEW INFRASTRUCTURE/$\Sigma$TPATH/0308/04 and the DECI resource Saniyer at UHeM
based in Turkey [http://en.uhem.itu.edu.tr] with support from the PRACE aisbl.
\end{acknowledgement}

{\bf Supporting Information Available:} The Supporting Information is available free of charge at
\url{link to be added by publisher}.\cite{Giannozzi_2017,Molina2013,Ponce2016a,Mostofi2008,Marzari2012,
pines1962,Caruso/PRB/2016,doi:10.1021/acs.jpcc.8b12029,doi:10.1021/acs.jpcc.5b07891,
PhysRevB.97.121201,Popescu_2012,Medeiros_2014,Zacharias_2020,jeong_spectroscopic_2019}
Supplementary Notes 1-6; Supplementary Figures S1-S12. 

\bibliography{references}

\end{document}